\documentclass[aps,pra,letterpaper,superscriptaddress,twocolumn,showpacs,floatfix,10pt]{revtex4-1}
\usepackage{amsfonts}
\usepackage{amsmath}
\usepackage{dsfont}
\usepackage{amssymb}
\usepackage{graphicx}
\usepackage{epstopdf}
\usepackage{epsfig}
\usepackage{times}
\usepackage{color}
\newcommand{\be}{\begin{equation}}
\newcommand{\ee}{\end{equation}}
\newcommand{\bea}{\begin{eqnarray}}
\newcommand{\eea}{\end{eqnarray}}
\newcommand{\ba}{\begin{array}}
\newcommand{\ea}{\end{array}}

\newcommand{\cH}{{\cal{H}}}

\newcommand{\om}{\omega}
\newcommand{\bt}{\beta}

\begin{document}
\title{Detection of Symmetry Enriched Topological Phases }
\author{Ching-Yu Huang}
\affiliation{Max-Planck-Institut f\"ur Physik komplexer Systeme, 01187 Dresden, Germany}
\author{Xie Chen}
\affiliation{Department of Physics, University of California, Berkeley, California 94720, USA}
\author{Frank Pollmann}
\affiliation{Max-Planck-Institut f\"ur Physik komplexer Systeme, 01187 Dresden, Germany}

\begin{abstract}

Topologically ordered systems in the presence of symmetries can exhibit new structures which are referred to as symmetry enriched topological (SET) phases. We introduce simple methods to detect the SET order directly from a complete set of topologically degenerate ground state wave functions. In particular, we first show how to directly determine the characteristic symmetry fractionalization of the quasiparticles from the reduced density matrix of the minimally entangled states. Second, we show how a simple generalization of a non-local order parameter can be measured to detect SETs. The usefulness of the proposed approached is demonstrated by examining two concrete model states which exhibit SET: (i) a spin-1 model on the honeycomb lattice and (ii) the resonating valence bond state on a kagome lattice. We conclude that the spin-1 model and the RVB state are in the same SET phases.
\end{abstract}

\maketitle

\paragraph*{Introduction.} Topologically ordered quantum systems have robust physical properties, like quasiparticle statistics and ground state degeneracy, which do not depend on the microscopic details of the Hamiltonian \cite{Wen1990}. If the system has extra global symmetries, then the interplay between topology and symmetry can give rise to interesting ``Symmetry Enriched Topological'' (SET) phases where the quasiparticles transform under the symmetry in a ``fractional'' way. The first and best understood topological phase -- the $\nu=1/3$ fractional quantum Hall state -- is an SET phase with charge conservation symmetry, where the quasiparticle with $e^{i\pi/3}$ fractional exchange statistics has $e/3$ fractional charge \cite{Tsui1982,Laughlin1983}. More interestingly, it was realized that systems with the same topological order and the same symmetry can be in different SET phases with different symmetry fractionalization on the quasiparticles. For example, in a $Z_2$ gauge theory with spin rotation symmetry, the gauge charges can carry half integer spin (fractional) or integer spin (non-fractional) representations \cite{Yao2010a}. With more symmetries, more varieties of SETs are possible and many efforts have been devoted to their classification \cite{Wen2002,Levin2012,Essin2013,Mesaros2013,Hung2013,Lu2013}.

An important open question is how to determine the SET order in a model system. With the experimental prospects to realize spin liquids in systems with various internal and lattice symmetries (e.g. herbertsmithite \cite{Han2012}), it is necessary to predict theoretically which SET phase they belongs to. However, this is generally hard as the SET order is intrinsically encoded in the global entanglement pattern of the state and no local order parameter can be measured to detect it. On the other hand, several methods have been developed to determine the topological order in the long-range entangled states \cite{Kitaev2006,Levin2006,Li2008,Tu2012,Zhang2012,Chen2010a}, but they are insensitive to the different ways of symmetry enrichment.

\begin{figure}[htbp]
\begin{center}
\includegraphics[width=8.0cm]{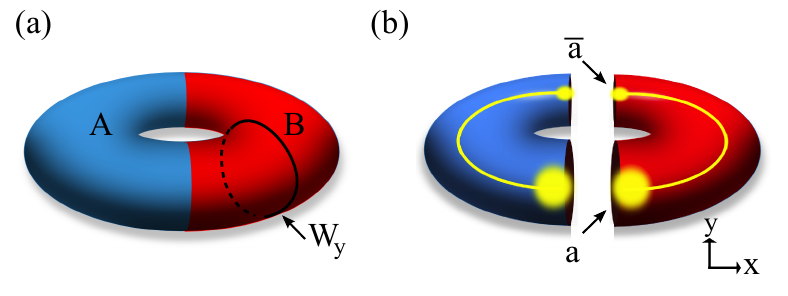}
\caption{(a) Minimally entangled ground states of a topological system on torus are eigenstates of Wilson loop $W_y$ operators parallel to the bipartite cut. (b) Quasiparticles of type $a$ and $\bar{a}$ are localized at the edges of the cut.}
\label{MES}
\end{center}
\end{figure}

In this paper, we introduce a way to detect SET order by measuring the non-local parameters on the minimally entangled ground states (MES) \cite{Zhang2012} of the system on a torus. The set of MES on a torus gives us access to the quasiparticle excitations of the system by localizing them at the ends of the cylinders when the torus is cut into halves, as shown in Fig.\ref{MES}. Now if we can measure the fractional symmetry representation carried by the quasiparticles, we can identify the SET order.  
%
If the quasiparticle carries a fractional charge, like in the fractional quantum Hall case, we can detect it directly by measuring charge locally near the ends of the cylinder in the MES. A different type of symmetry fractionalization exists where the quasiparticles carry projective representations (see Appendix \ref{app_a}) of the symmetry, as in the case of spin $1/2$ representations on the $Z_2$ gauge charges. We are going to focus on SETs with this type of fractionalization in this paper and show that they can be determined with the non-local order parameters that are related to the ones used to detect symmetry protected topological phases (SPT) \cite{Pollmann2012a,Haegeman2012}.
As shown in Refs. \cite{Pollmann2012a,Haegeman2012}, a string order parameter can be designed to detect projective symmetry representations on the edge of a one-dimensional (1D) gapped system hence identifying the SPT order in 1D.
By putting a two-dimensional (2D) SET system onto a cylinder and picking out the MESs, we map a 2D SET state into effectively 1D SPT states whose order can then be detected with ``non-local order'' parameters.
We demonstrate the effectiveness of this idea by applying it to a model wave function of spin-1 bosons and also to the resonating valence bond state \cite{ANDERSON1987,Baskaran1987} on the kagome lattice which has the same SET order -- the $Z_2$ topological order with the $Z_2$ charge carrying spin $1/2$ representation of the $SO(3)$ spin rotation symmetry.

\begin{figure}[ht]
\center{\epsfig{figure=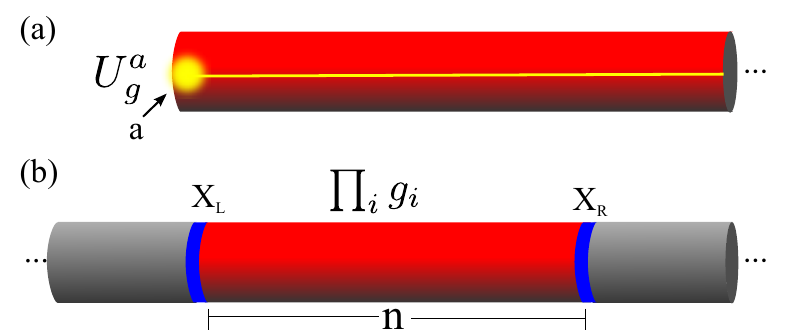,angle=0,width=9cm}}
\caption{(a) Symmetry fractionalization for a localized quasiparticle of type $a$ at the end of the half cylinder. (b) The non-local order parameter consisting of product of onsite symmetry operators $\Sigma$  acting on a segment of length $n$ on the cylinder. The segment is terminated by the operators $X_L$ and $X_R$ .\label{string}}
\end{figure}


\paragraph*{Detecting SET's.}

An important concept for the detection of SETs are the minimally entangled states proposed in \cite{Zhang2012}.
When a topologically ordered system is put onto a torus, the ground space has a degeneracy equal to the number of quasiparticle types $N$.
Among all the states in the ground subspace, one set of basis states have minimum bipartite entanglement when the torus is cut along a non-contractible loop in the $y$ direction into two cylinders as shown in Fig.~\ref{MES}.
These minimally entangled states (MES) are eigenstates of the Wilson loop operators in the $y$ direction $\{W^y_{a}\}$ and have a one to one correspondence with the quasiparticle types $a$.
The MES with eigenvalue $1$ for all $\{W^y_{a}\}$ corresponds to the vacuum. MES corresponding to $a$ can be created from the vacuum state by creating a pair of particle $a$ and anti-particle $\bar{a}$ excitations, bringing $a$ around a non-contractible loop in the $x$ direction and annihilating it with $\bar{a}$.
Therefore, when the system is cut open, the MES corresponding to $a$ has quasiparticles $a$ and $\bar{a}$ at each end of the cylinder.
Given a model Hamiltonian with potential SET order, one can find the MESs from a complete set of ground states on a torus by minimizing bipartite entanglement. In the following, we follow Refs.~\cite{cincio2013, Zaletel2013} and use the fact that a long cylinder is locally equivalent to a torus.
In this case, the different MESs can be conveniently obtained by changing the boundary conditions on a long cylinder.
When cut into two ``half cylinders'',  we find a localized quasiparticle near the edges as illustrated in Fig.~\ref{string}a with possible projective representations.
A projective representation is like an ordinary representation up to phase factors; i.e., if $g,h,k$ are in $G$ and fullfill  $g\cdot h=k$,
then
\begin{equation}
U_{g}U_{h}=e^{i\rho(g,h)}U_{k}.
\label{eq:rho}
\end{equation}
The phases $\rho(g,h)$ are called the ``factor set" of the representation.
In Ref.~\onlinecite{Pollmann2012a}, a  numerical approach was introduced that allows to directly extract the $U_g$ from the ground-state wave functions.
Furthermore, several string order parameter were introduced which allow to directly detect different SPT phases, including those protected by time reversal and by inversion symmetry.

We use two methods to identify the SET phase from the MESs in this paper.
We begin with the method in which the projective representations $U_g$'s are directly extracted from the Schmidt decomposition of the MES. Assume that $|\Xi^a\rangle$ is the MES corresponding to quasiparticle $a$ on an infinite cylinder and perform a Schmidt decomposition of the state into two Schmidt states on half cylinders
\begin{align}
|\Xi^a\rangle=\sum_{\alpha=1}^\chi \lambda_{\alpha} |\phi_{\alpha,a}^L\rangle |\phi^R_{\alpha,a}\rangle,
\end{align}
where $ |\phi_{\alpha,a}^L\rangle$ and $ |\phi_{\alpha,a}^R\rangle$ represent an orthogonal basis of the left and right partitions, respectively.
For concreteness, we assume that the $ |\phi_{\alpha,a}^L\rangle$ are defined on sites $-\infty\dots 0$ and  $ |\phi_{\alpha,a}^R\rangle$ on sites $1\dots\infty$.
The $\lambda_\alpha$'s are Schmidt values and the entanglement entropy is given by $S=-\sum_\alpha \lambda_\alpha^2 \log \lambda_\alpha^2$.
Note that the Schmidt states can also be obtained by diagonalizing the reduced density matrix $\rho^L$ ($\rho^R$) and the corresponding eigenvalues are $\lambda_{\alpha}^2$.
The SETs are gapped phases with short range correlations and we assume that we have a cylinder with a finite circumference, thus the area law \cite{Srednick1993} guarantees that the values of $\lambda_{\alpha}$ decay quickly  \cite{Gottesman-2009, Hastings-2007}. From now on,  we are only considering the \emph{important} Schmidt states which have a Schmidt value $\lambda_{\alpha}>\epsilon$ for a given $\epsilon>0$, so we have a finite number of Schmidt values.

The Schmidt states of $|\Xi^a\rangle$ have localized quasiparticles of type $a$ at the cut.
Thus onsite symmetry operations $g$ that act on the Schmidt states transform the quasiparticles according to the representation $U_g^a$ (which can be either linear or projective).
The $U_g^a$ can then be directly obtained by calculating the overlap between the Schmidt states with their symmetry transformed partners,
\begin{align}
(U_g^a)_{\alpha,\beta}=  \langle \phi_{\alpha,a}^R| \left( \prod_{i=1}^{\infty} g_i \right)  |\phi_{\beta,a}^R\rangle.\label{eq:getu}
\end{align}
Equivalently, we could have chosen the left Schmidt states $|\phi_{\alpha,a}^L\rangle$.
If the state $|\psi_0\rangle$ is represented as tensor product state, Eq.~(\ref{eq:getu}) can be efficiently evaluated  by multiplying together all the tensors to the right of the bond (see Appendix \ref{app findU}).
Once we have obtained the $U^a_g$ of each symmetry operation and quasiparticle type, we can calculate the commutators and read off the factor set and hence determine in which phase the state is.
For non-onsite symmetries, we can obtain the representations in an analogous way, e.g., for time-reversal  and inversion symmetry.

%
It is desirable to have a probe for SETs that does not rely on having access to the Schmidt states.
For this we construct non-local order parameters that are sensitive to the type of SET order and can be directly evaluated using Monte Carlo methods or potentially measured in experiments \cite{Endres2011}.
The non-local order parameter we consider here is defined as
\begin{eqnarray}
&\mathcal{O}^{a}&(g,X^L,X^R)=\nonumber\\
&\lim&_{n\rightarrow\infty}\left\langle \Xi^a\left| X^L(1)\left(\prod_{k=2}^{n-1}g(k)\right)X^R(n)\right|\Xi^a\right\rangle.\label{eq:so}
\end{eqnarray}
and closely related to the string order parameter originally introduced by \cite{Nijs1989}.
It corresponds to calculating the overlap between the wave function with a symmetry operation $g$ applied to a segment of $n$ consecutive rings of the cylinder (as illustrated in Fig.~\ref{string}b).
The operators $X^L$ and $X^R$ are defined on rings terminating the segment that is transformed by $g$.
As $g$ is a symmetry operation, it does not change anything in the bulk of this segment and the overlap should not vanish as $n\rightarrow \infty$ for any cylinder with a finite circumference.
In Ref.~\cite{Pollmann2012a}, it was demonstrated, that string order parameter of the type Eq.~(\ref{eq:so}) can detect SPT phases by choosing the operators $X^L$ and $X^R$ accordingly.
Let us assume that the quasiparticles of type $a$ transform as $U^a_g U^a_h = e^{i\rho(g,h)}U^a_k$.
The operator $X (= X^L = X^R)$ can be chosen to have a particular quantum number $e^{i\sigma(X, h)}$ with respect to $h$.
This yields a selection rule and the non-local order parameter has to be zero if $\rho(g,h) \ne  \sigma(X, h)$.
In practice, we can thus choose the operators $X^L$ and $X^R$ accordingly and then evaluate Eq.~(\ref{eq:so}) for each quasiparticle type to get a complete characterization of the SET.
Note, that the non-local order parameter (see Eq.~(\ref{eq:so})) is sensitive for a subset of SETs with sufficiently simple symmetries, however, more general order parameters can be constructed that work for all symmetries (analogous to the ones used for SPT in \cite{Pollmann2012a}).

\paragraph*{Spin-1 Bosons on the hexagonal lattice.}
We will now demonstrate the above by studying an example of an SET that is built of spin-1 bosons.
In particular, we are going to construct a simple SET state, namely, an ``Affleck, Kennedy, Lieb and Tasaki (AKLT) string model state'' and show how we can then extract all characteristic properties.
The state we consider here is defined on a honeycomb lattice where each site is either unoccupied or contains one spin-1 boson. The ground state wave function is an equal weighted superposition of loop coverings on the honeycomb lattice, where along the loops the spin-1 bosons form AKLT chains \cite{Affleck1988} and away from the loops the sites are unoccupied as shown in Fig.~\ref{spin-1_boson}a. In Appendix \ref{H_spin1boson}, we describe a local Hamiltonian which has the state as its ground state.
Neglecting the internal structure of the loops, this state is exactly the ground state of the toric code model \cite{Kitaev2003} -- which is the fixed point  of a $\mathbb{Z}_2$ topologically ordered phase \cite{Chen2010,LevinGu2012}.
The topological entanglement entropy (TEE), usually denoted by $\gamma$, is the constant term in the entanglement entropy $S=cL-\gamma$,  where $L$ is the length of the boundary of the region\cite{Levin2006, Kitaev2006}.
The  characteristic topological entanglement entropy of the $\mathbb{Z}_2$ phase is $\gamma=\log 2$.
The $\mathbb{Z}_2$ phase has a four-fold ground state degeneracy on a torus, which corresponds to four different types of quasiparticles excitations.
These are the electric particles $e$ (showing up as the ends of open strings in Fig.~\ref{spin-1_boson}b), the magnetic particles $m$ (corresponding to defects on  plaquettes), fermions $f$ (bound pairs of $e$ and $m$), and the identity particle $1$.
The four quasiparticle types are related to the $\pm1$ eigenspaces of $\{W^y_{a}\}$.

In the AKLT string model,  the sites have integer spin $S=0$ or $S=1$. However, the $e$ particles appear at ends of open AKLT loops and therefore carry a fractionalized spin $S=1/2$ as shown in Fig.~\ref{spin-1_boson}b.
Let us now assume the presence of a symmetry, e.g., $SO(3)$,  a discrete subgroup like $\mathbb{Z}_2 \times \mathbb{Z}_2$ or time reversal symmetry.
We find that the onsite representations are always linear, while the half integer spin carried by the $e$ particle has a projective representation. The $m$ particle carries integer spin and their bound state $f$ has half integer spin.
%
%

The AKLT string state can be represented exactly by a tensor product state (TPS) \cite{Verstraete2004} which simplifies the calculations considerably.
The state with $n$ sites can be expressed as a (translationally invariant) TPS with bond dimension $\chi=3$ as follows:
\begin{align}
|\Psi\rangle=\sum_{s_1,s_2,...,s_n} \text{tTr}[T^{[s_1]} T^{[s_2]}... T^{[s_n]} ] |s_1,s_2,...,s_n \rangle,
\end{align}
where the $s_i\in \{1,0,-1,\varnothing \}$ correspond to the three $S=1$ states  and the vacuum state, respectively.
The non-zero elements of the tensors are
\begin{align}\label{z2TPS}
&T_{222}^{[\varnothing]}=1; \\ \notag
&T_{201}^{[0]}=T_{021}^{[0]}=T_{012}^{[0]}= -T_{210}^{[0]}=-T_{120}^{[0]}=-T_{102}^{[0]}= \frac{t}{\sqrt{2}}; \\ \notag
&T_{112}^{[-1]}=T_{121}^{[-1]}=T_{211}^{[-1]}=-T_{002}^{[1]}=-T_{020}^{[1]}=-T_{200}^{[1]}=-t,
\end{align}
where $t$ is a string tension. If $t=1$, this state is an equal weighted superposition of all ``AKLT loop'' coverings  (see Appendix \ref{app_SET_tensor} for the details of the derivation). Starting from this set of tensors, we can find the TPS representation for all the MES (for details see Appendix \ref{app findU}).

As described in Refs.~\cite{Cirac2011,Schuch2013}, we can obtain the reduced density matrix for a TPS on the cylinder of circumference $L$ for each MESs (the details are outlined in Appendix \ref{app findU}).
In particular, the Schmidt states $|\phi^L_{\alpha,a}\rangle$ of each MES can be obtained by diagonalizing the reduced density matrix on the cylinder.
From the entanglement entropy, which scales with $L$ as $cL-\gamma$, we can then directly obtain the constant $\gamma$.
We obtain for  all MES a $\gamma$ which converges to $\log 2$ as $L$ increases (see Fig \ref{spin-1_boson}d).
This is the expected result for a $\mathds{Z}_2$ topologically ordered phase.

\begin{figure}[ht]
\center{\epsfig{figure=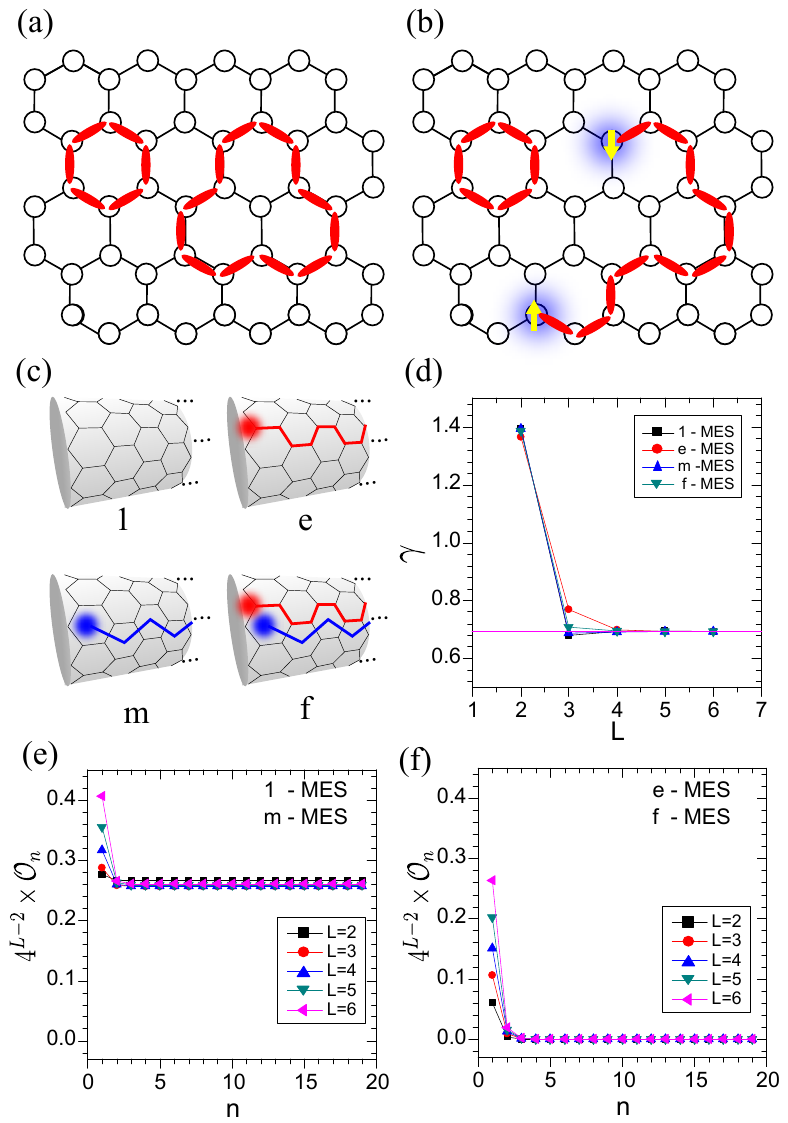,angle=0,width=8cm}}
\caption{(a) The ground state wave function is formed by an equal weighted superposition of closed $S=1$ AKLT chains on the honeycomb. The sites are are either occupied by one $S=1$ boson or empty and the red ellipsoids represent spin 1/2 singlets. (b) An excited state with two defects which carry a spin 1/2 each. (c) The four MES of the $\mathds{Z}_2$ liquid are in a one-to-one correspondence with  the four quasiparticle types $1,e,m$ and $f$. (d) All MES have a topological entanglement of $\gamma=\log 2$. (e-f) Non-local order parameter shown for a string tension of $t=0.83$ with identity as boundary operator ($X=\mathds{1}$ shown in Eq.~(\ref{eq:so}) ) as a function of the length of the segment $n$ calculated for the four MES for cylinders of different circumference $L$. $\mathcal{O}_n$ decays exponentially with $L$ and we have rescaled the quantity in the plot.\label{spin-1_boson}}
\end{figure}

By inserting the symmetry operators, such as time reversal operator or the $\pi$ rotation operators ($R^x=\exp(i\pi S^x)$ and $R^z=\exp(i\pi S^z)$) into the transfer matrix, we directly obtain the overlap Eq.~(\ref{eq:getu}) yielding the desired $U^a_g$.
From the $U^a_g$, we can then calculate the commutators which characterize the SET:
\begin{center}
\begin{tabular}{l c c c c r }
\hline  & 1& $e$ & $m$ & $f$ \\
\hline  $U^a_{TR}(U^a_{TR})^*$ &  1.0 &  -1.0 & 1.0  &  -1.0 \\
 $U^a_{R^x}U^a_{R^z}(U^a_{R^x})^{\dag}(U^a_{R^z})^{\dag}$  &  1.0 &  -1.0 & 1.0  &  -1.0  \\
 \hline
\end{tabular}
\end{center}
We find that the commutators of $U^a_g$ reveal nontrivial phase factors for the time reversal symmetry and $\mathbb{Z}_2\times\mathbb{Z}_2$ symmetry in the $e-$ and $f-$MES.
This is the fingerprint of the specific SET.
We also demonstrate how  to detect SET phases by using non-local order parameter Eq.~(\ref{eq:so}). The non-local order parameter with $X=\mathds{1}$ and $g=R^x$ shown in Figs. \ref{spin-1_boson}e and \ref{spin-1_boson}f reveals the projective representations of the $e$ and $f$  quasiparticles.
The selection rule implies that the non-local order parameter $\mathcal{O}_n(R^z,\mathds{1},\mathds{1})$ vanishes in $e-$ and $f-$MES as they have non-trivial phase factors under $R^z$.
%

\paragraph*{Resonating valence bond state on the kagome lattice.}
\begin{figure}[ht]
\center{\epsfig{figure=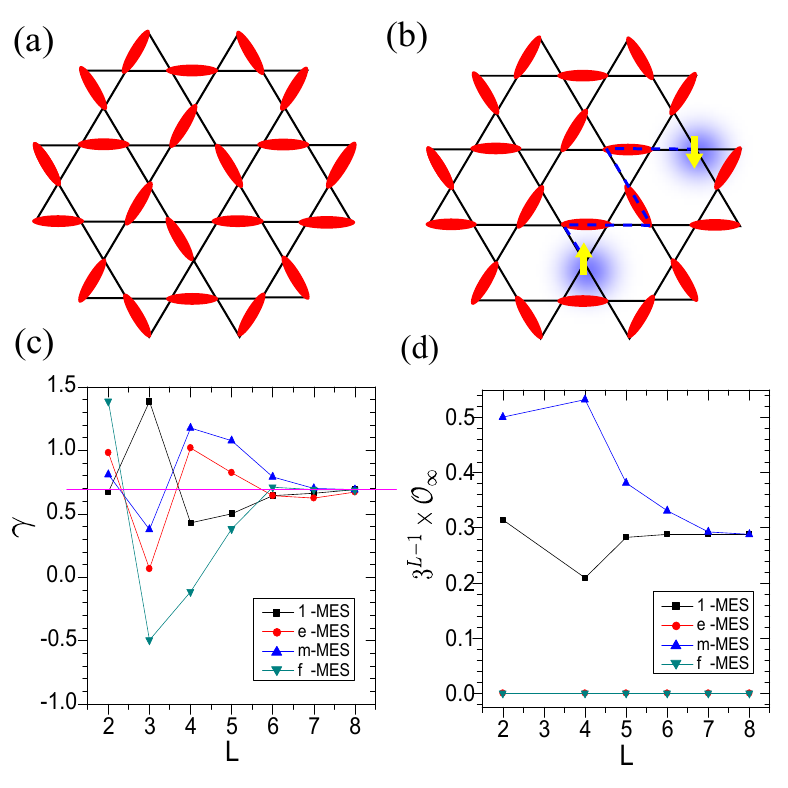,angle=0,width=8cm}}
\caption{(a) The RVB state on the kagome lattice is given by an equal weighted superposition of nearest-neighbor singlet coverings. The red ellipsoids represent spin 1/2 singlets. (b) A pair of spinon excitations carrying a spin 1/2 each. (c) All MES have a topological entanglement of $\gamma=\log 2$. (d) Non-local order parameter for each MES.\label{RVB}}
\end{figure}

We will now show that the resonating valence bond state (RVB) state on the kagome lattice (shown in Fig. \ref{RVB}a) is in the same SET phase as the spin-1 model above.
The RVB state on the kagome lattice is a good approximation of the ground state for the $S=1/2$ Heisenberg \cite{Poilblanc2010} and represents a state with $\mathbb{Z}_2$ topological order \cite{Schuch2012}.
In the RVB state, a singlet can fractionalize into two spinons which carry a  spin $1/2$ (see Fig. \ref{RVB}b).
Thus the quasiparticles  again carry  projective representations.
Again we can make use of the TPS description of the state with a bond dimension $\chi=3$ as shown in Ref.~\onlinecite{Schuch2012} and the calculations of the non-local order parameters can be performed analogously.
We first obtain the TEE of each of the four MESs that tend to $\log 2$ for large $L$ as shown in Fig. \ref{RVB}c -- as expected for a $\mathds{Z}_2$ phase.
The projective representations $U^a_g$ are the same as result in the previous example.
The non-local order parameter vanishes in $e-$ and $f-$MES and tends to a constant in $1-$ and $m-$MES as shown in Fig. \ref{RVB}d.
From the TEE and non-local parameter, we can conclude that the RVB state is in the same phase as the AKLT string model.

\paragraph*{Conclusions.}
In this paper, we have introduced two simple methods to detect SET.
The first approach is achieved by the Schmidt states of minimally entangled state (MES) on a cylinder.
We can measure the symmetry representations of quasiparticle type $a$ at the end of cylinder.
Thus it is a very convenient method if we use exact diagonalization (see Appendix \ref{app_ED}), density matrix renormalization group  \cite{Whit1992}, or tensor product state based \cite{Cirac2011} techniques.
The second approach is to do a segment of  measurements on the real spins. The selection rules obtain a characterization of SET.
This way is more physical, and can be used by other methods, e.g., quantum Monte Carlo methods or potentially measured experimentally.
We demonstrated the usefulness of this approach by considering first a AKLT string model and the RVB state on the kagome lattice and find that they are in the same SET phase.
%

\paragraph*{Acknowledgment.}
XC is supported by the Miller Institute for Basic
Research in Science at UC Berkeley. We are grateful for discussions with Ashvin Vishwanath, Ari Turner, Mike Hermele, Mike Zaletel, Norbert Schuch, and Siddhardh Morampudi.

\bibliographystyle{apsrev_nurl}	
\bibliography{bibs}

\appendix
\section{Projective Representation}\label{app_a}

Matrices $u(g)$ form a projective representation of symmetry group $G$ if
\begin{align}
 u(g_1)u(g_2)=\om(g_1,g_2)u(g_1g_2),\ \ \ \ \
g_1,g_2\in G.
\end{align}
Here $\om(g_1,g_2)$'s are $U(1)$ phase factors, which is called the
factor system of the projective representation. The factor system satisfies
\begin{align}
 \om(g_2,g_3)\om(g_1,g_2g_3)&=
 \om(g_1,g_2)\om(g_1g_2,g_3),
\end{align}
for all $g_1,g_2,g_3\in G$.
If $\om(g_1,g_2)=1$, this reduces to the usual linear representation of $G$.

A different choice of pre-factor for the representation matrices
$u'(g)= \bt(g) u(g)$ will lead to a different factor system
$\om'(g_1,g_2)$:
\begin{align}
\label{omom}
 \om'(g_1,g_2) =
\frac{\bt(g_1g_2)}{\bt(g_1)\bt(g_2)}
 \om(g_1,g_2).
\end{align}
We regard $u'(g)$ and $u(g)$ that differ only by a prefactor as equivalent
projective representations and the corresponding factor systems $\om'(g_1,g_2)$
and $\om(g_1,g_2)$ as belonging to the same class $\om$.

Suppose that we have one projective representation $u_1(g)$ with factor system
$\om_1(g_1,g_2)$ of class $\om_1$ and another $u_2(g)$ with factor system
$\om_2(g_1,g_2)$ of class $\om_2$, obviously $u_1(g)\otimes u_2(g)$ is a
projective presentation with factor group $\om_1(g_1,g_2)\om_2(g_1,g_2)$. The
corresponding class $\om$ can be written as a sum $\om_1+\om_2$. Under such an
addition rule, the equivalence classes of factor systems form an Abelian group,
which is called the second cohomology group of $G$ and denoted as
$\cH^2[G,U(1)]$.  The identity element $1 \in \cH^2[G,U(1)]$ is the class that
corresponds to the linear representation of the group.

\section{Hamiltonian for the AKLT string model} \label{H_spin1boson}

\begin{figure}[htbp]
\center{\epsfig{figure=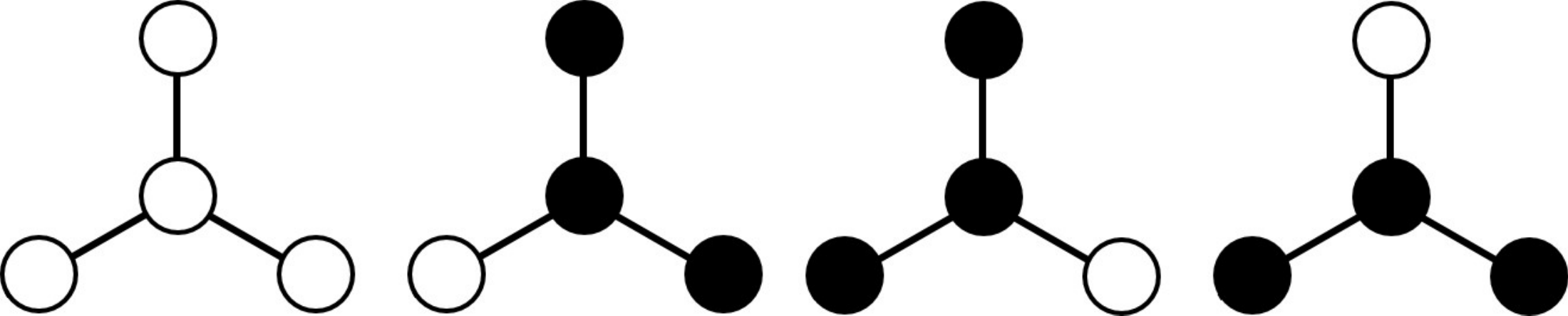,angle=0,width=8cm}}
\caption{Vertex configuration allowed by the $h_v$ term. Empty circle represents the vacuum sector while solid circle represents the spin-1 boson sector.}
\label{h_v}
\end{figure}

\begin{figure}[htbp]
\center{\epsfig{figure=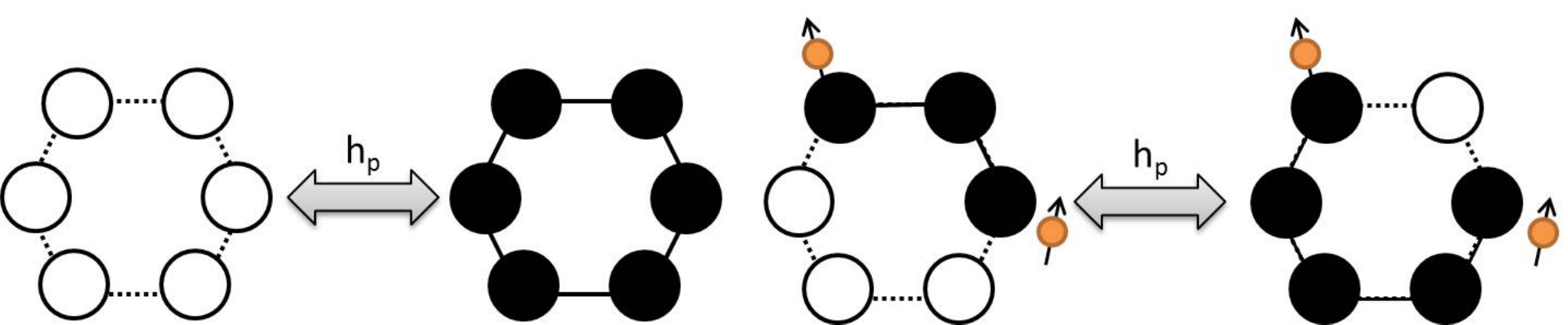,angle=0,width=8cm}}
\caption{Plaquette operator allows Haldane loops to fluctuate. Small spins on the side indicate edge spins of the Haldane chain segment.}
\label{h_p}
\end{figure}

In this section, we describe a Hamiltonian which has the spin-1 boson wave function -- an equal weighted superposition of AKLT loops -- as its ground state. The Hamiltonian contains a vertex term and a plaquette term.
\be
H=\sum_v h_v + \sum_p h_p
\ee
Each $h_v$ acts on four vertices on the hexagonal lattice, with one vertex in the center and three neighboring vertices around it. First $h_v$ projects onto the following subspace: if the center vertex is in the vacuum sector, then all three neighboring vertices are in the vacuum sector; if the center vertex is in the one boson sector, then two of the neighboring vertices are also in the one boson sector while the other one is in the vacuum sector, as shown in Fig~\ref{h_v}.
Moreover, the vertex term contains a coupling term $\vec{S}_i \cdot \vec{S}_j+\frac{1}{3}\left(\vec{S}_i \cdot \vec{S}_j\right)^2$ between the spin-1 at the center and each of the two spin-1's at the neighboring sites. The low energy space of all the vertex terms is then composed of loop configurations of AKLT chains. The plaquette term $h_p$ then allows the AKLT loop configurations to fluctuation from one to another. More specifically, if a plaquette does not have a loop initially, the plaquette operator would attach an AKLT loop to it, as shown in Fig.\ref{h_p}. If a loop overlaps with part of the plaquette, then the plaquette term would flip the loop to be on the other side of the plaquette. Note that a segment of an AKLT chain has four low energy states associated with the two edge spin $1/2$'s. When mapping the segment from one side of the plaquette to another side, we need to match the state of the edge spin.

\section{Tensor product states for the AKLT string model and the RVB state}\label{app_SET_tensor}

In this appendix, we explain the AKLT string model on the honeycomb lattice and the RVB state on the kagome lattice in TPS formalism.

\subsection{TPS representation of the AKLT string model}
Let us first recall the toric code  model \cite{Kitaev2003} for which the ground state is the fixed point of a $\mathbb{Z}_2$ topologically ordered phase.
The model is defined on the honeycomb lattice where each link is occupied by an Ising spin.
The Hamiltonian is given by the spins around each vertex $v$, $-\sum_v \prod_{i\in v}Z_i$  and plaquette $p$  of the lattice, $-\sum_p \prod_{i\in p}X_i$.  Here, $X$ and $Z$ are Pauli operators.
If we define that the state $|\uparrow\rangle$ of spin corresponds to a string on a link, the ground state wave function is an equal weighted superposition of all closed loop configurations  $|\Psi_{\mathbb{Z}_2}\rangle=\sum |D\rangle$  on the lattice.

Now we start from a state, $|D_s\rangle$, which is a tensor product of singlets $|\uparrow\downarrow\rangle-|\downarrow\uparrow\rangle$ between two connected sites in the loop covering  $|D\rangle$.
The state $|D_s\rangle$ can be regarded as a covering of the honeycomb lattice with 1D $S=1$ AKLT chains. Each site is occupied by either two spin $1/2$ (that have a total spin  $S=1$) or a vacuum state (no string crosses this site).
The equal weighted superposition of all AKLT loop coverings, $|\Psi\rangle=\sum |D_s\rangle$, on the honeycomb lattice forms the ``AKLT string" model state.
This state has  $\mathds{Z}_2$ topological order and is a simple example of an SET (as described in the main text).

To obtain a TPS representation of the AKLT string model, we can place the entangled state $|\omega\rangle = \frac{1}{\sqrt{2}}( |01\rangle-|10\rangle)+|22\rangle$ along all edges of the lattice.
Each physical site has three virtual particles  on the honeycomb lattice, and  each  of virtual particles can be regarded as spin $1/2\oplus 0 $.
The virtual indices ``0,1'' provide the spin $1/2$ degrees of freedom which holds singlet along the edge.
The third index ``2" that belong spin $0$ subspace is used to indicate there is no singlet along the edge.
At each physical site, if two of the virtual particles stay in spin $1/2$ subspace and the other one stays in spin $0$, these three virtual particles would be mapped to a physical spin-1 boson $(S=1,0,-1)$.
If all of the virtual particles stay in spin $0$, it forms a physical vacuum state, $\varnothing$.
The following projector realizes the spin-1 boson on the honeycomb lattice,
\begin{align}
P=&t|1\rangle ( \langle 002|+\langle 020|+\langle 200|)+\\ \notag
  &t|0\rangle (\langle 012|+\langle 120|+\langle 201|+\langle 021|+\langle 210|+\langle 102|)+\\ \notag
  &t|-1\rangle (\langle 112|+\langle 121|+\langle 211|)+\\ \notag
  &|\varnothing\rangle \langle 222|,
\end{align}
here, $t$ is a string tension. Each projector $P$ on site with $t=1$ represents an equal weighted superposition of all AKLT loop coverings. By contracting the virtual particles with projectors $P$, we can obtain the TPS representations directly.

\subsection{TPS representation of the RVB state}\label{app_RVB_tensor}

The tensor representation of the RVB state also can be obtained by using $\chi=3$ entangled state. In Ref.~\cite{Schuch2013},  a 3-particle entangled state, $|\epsilon \rangle=\sum_{i,j,k=0}^2 \varepsilon_{ijk}|ijk\rangle+|222\rangle$ was used, and placed on a triangle of kagome lattice. The $\varepsilon$ is the Levi-Civita symbol with $\varepsilon_{012}=\varepsilon_{120}=\varepsilon_{201}=1$ and $ \varepsilon_{021}=\varepsilon_{210}=\varepsilon_{102}=-1$. Applying the projector
\begin{align}
P=&|\uparrow\rangle (\langle 02|+\langle 02|)+|\downarrow\rangle(\langle 12|+\langle 21|)
\end{align}
to each vertex yields the desired RVB state.

\section{Details on how to obtain the projective representations of the anyons in the $\mathds{Z}_2$ model from a TPS}\label{app findU}

In this appendix, we show the details of how to obtain the projective representations $U^a_g$ of the anyons from TPS representations.
The main procedure is now as follows:
we firstly obtain TPS representations of the anyons from a complete set of the ground state.
We then map a 2D TPS to an effective 1D matrix product state (MPS) representation $\tilde{A}$, and find the canonical form of the matrix $\tilde{A}$.
Finally, we can obtain the projective representations $U_g^a$ of the anyons from the symmetry transformations.

Let us consider a wave function, $|\Psi\rangle$, of an $N_1\times N_2$ spin lattice in a cylindrical geometry. It can be expressed in terms of a TPS as
\begin{align}\label{TPSsim}
|\psi\rangle = \sum_{s_{1,1},s_{1,2},\ldots,s_{N_1,N_2}}
&\text{tTr} (T^{s_{1,1}}T^{s_{1,2}}\ldots T^{s_{N_1,N_2}})\\ \notag
&|s_{1,1}s_{1,2}\ldots s_{N_1,N_2}\rangle,
\end{align}
where the physical spin $s_{i,j}=1,...,d_s$, and $T^{s_{i,j}}$'s are rank-five tensors for a square lattice. In the AKLT string model we considered, the honeycomb lattice also can be  mapped onto an effective square lattice (see Fig.\ref{TPS}a).

Our proposed method is implemented as follows.

\begin{enumerate}
  \item Find the TPS representations of MESs:

  We reiterate how the TPS representation of the minimally entangled states corresponding to the anyons \cite{Schuch2013} can be obtained.
  In $\mathbb{Z}_2$ topologically ordered phases, the TPS representations of four anyons can be obtained directly from the complete set of ground states.
  Suppose that we have the ground states of a $\mathbb{Z}_2$ topologically ordered phase in TPS corresponding to Eq.~(\ref{TPSsim}).
  The MESs of quasiparticles $1$ and  $e$, which are related to the $\pm1$ eigenspaces of $W_a^y$, can be obtained from the TPS for the ground state with even and odd parity number of the boundaries of a cylinder.
  The parity number is defined by counting the number of singlets that cross a vertical line of the cylinder.
  These two MESs can be written as simple TPS representations.

  We also can insert a string operator $\bar{Z}\otimes\bar{Z}....\otimes\bar{Z}$ (as the blue line shown in Fig. \ref{spin-1_boson}c ) to TPSs for $1-$ and $e-$MES to create magnetic fluxes. The representation of the  operator $\bar{Z}$ is given by
  \begin{align}
  \bar{Z}=\left(
  \begin{array}{ccc}
        1 & 0 & 0 \\
        0 & 1 & 0 \\
        0 & 0 & -1 \\
      \end{array}
    \right).
    \end{align}
  Again, we can obtain TPS representations for $m-$ and $f-$MES with string operator.

  \item  Map 2D TPS to 1D MPS:

   We consider a MES corresponding to quasiparticle $a$ in TPS Eq.~(\ref{TPSsim}) (here the index of the anyon is omitted).
   Following  Refs.~\cite{Cirac2011,Schuch2013}, we block all spins that are in the first $k$ columns to form a ring, and define a new tensor,
  \begin{align}
  \tilde{ A}_{\hat{\alpha_k},\hat{\gamma_k} }^{k,S_k}  = \sum_{\beta_{1,k},...,\delta_{N_1,k}}  T^{s_{1,k}}T^{s_{2,k}}...T^{s_{N_1,k}},
  \end{align}
  as shown in Fig. \ref{TPS}b. Here, $S_k$ denotes the combination of all physical indices $s_{1,k},s_{2,k},...,s_{N_1,k} $, and $\hat{\alpha_k}$ and $\hat{\gamma_k}$ denote the combination of all inner indices $\alpha_{1,k},\alpha_{2,k},...,\alpha_{N_1,k} $ and $\gamma_{1,k},\gamma_{2,k},...,\gamma_{N_1,k} $ respectively.
  In terms of those tensors $\tilde{A}$'s, the MES can be expressed as
  \begin{align}
  |\psi\rangle = \sum_{S_1S_2...S_{N_2}}
  \text{Tr} &(\tilde{A}^{1,S_{1}}\tilde{A}^{2,S_{2}}...\tilde{A}^{N_2,S_{N_2}})\\ \notag
  &|S_1S_2...S_{N_2}\rangle,
  \end{align}
  where $S_i$ is a physical index of a new tensor which includes all physical indices around a ring.
  A 2D TPS thus can be mapped to an effective 1D MPS with large physical and inner dimensions shown in Fig. \ref{TPS}c.

  \begin{figure}[ht]
  \center{\epsfig{figure=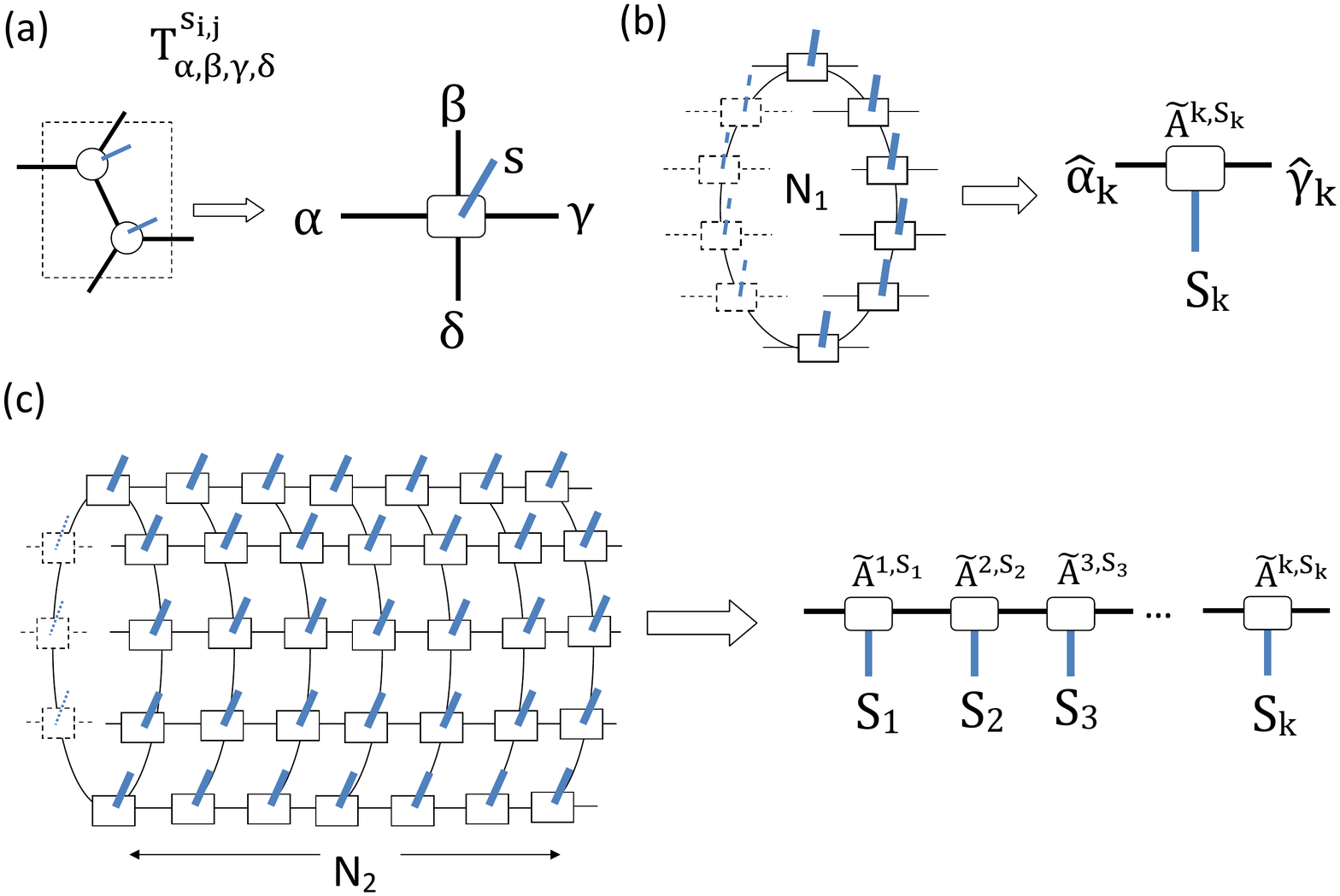,angle=0,width=8cm}}
  \caption[TPS]
  {(a) Tensors are denoted as boxes with one physical index $s$ and four virtual indices $\alpha,\beta,\gamma,\delta$.  (b) meet all of spins as a ring and form a new tensor $\tilde{A}$. (c) The effective MPS can be obtained by replacing  the tensors around a ring with a new tensor $\tilde{A}$ and contracting the virtual indexes along the horizontal direction.  }
   \label{TPS}
   \end{figure}

  \item  Determine the canonical form:

  Now, we have an effective 1D MPS and need to determine the canonical form \cite{Orus2008} of the same state.
  The procedure is covered in detail.
  First, we form a positive double tensor $\mathbb{E}$ by merging two layers of tensor $\tilde{A}$ and $\tilde{A}^*$ with the physical indices contracted, namely,
  \begin{align}
  \mathbb{E}_{\hat{\alpha}\hat{\alpha}' ,\hat{\beta}\hat{\beta}'}=\sum_{S} \tilde{A}^{* S}_{\hat{\alpha}',\hat{\beta}'} \tilde{A}^S_{\hat{\alpha},\hat{\beta}}.
  \end{align}
  Find the $V_R$ that is the dominant right eigenvector of double tensor $\mathbb{E}$ with eigenvalue $\eta$ (see the Fig. \ref{canonical}a). We prepare an initial vector $V_{i}$  and apply the double tensors $N$ times.
  By using power method, in the limit $N\rightarrow\infty$,  $( \mathbb{E}_{\hat{\alpha}\hat{\alpha}' ,\hat{\beta}\hat{\beta}'} ) ^N (V_i)_{\hat{\beta}\hat{\beta}'}$ will converge to the dominant right eigenvector of $\mathbb{E}$.
  Here, we normalize $V_R$ such that $V_RV_R^\dag$ = 1 and ignore a constant phase factor which results from the right end.
  Then, decompose the matrix $(V_R)_{\hat{\alpha}\hat{\alpha}'}$, which is Hermitian and non-negative, as $V_R=W\sqrt{\lambda}\sqrt{\lambda}W^\dag=XX^\dag$.
  Finally, we arrange the tensors $\tilde{A}$ and $X$ into a new tensor
  \begin{align}
  \tilde{A}^{'S}_{\hat{\alpha},\hat{\beta}}=\sum_{\hat{\gamma},\hat{\delta}}X^{-1}_{\hat{\alpha},\hat{\gamma}}\tilde{A}^S_{\hat{\gamma},\hat{\delta}}  X_{\hat{\delta},\hat{\beta}}.
  \end{align}
  The tensor $\tilde{A}'$ is thus in the canonical form that defined by the right eigenvector of $\mathbb{E}$ as shown in Fig. \ref{canonical}b.

  \begin{figure}[ht]
  \center{\epsfig{figure=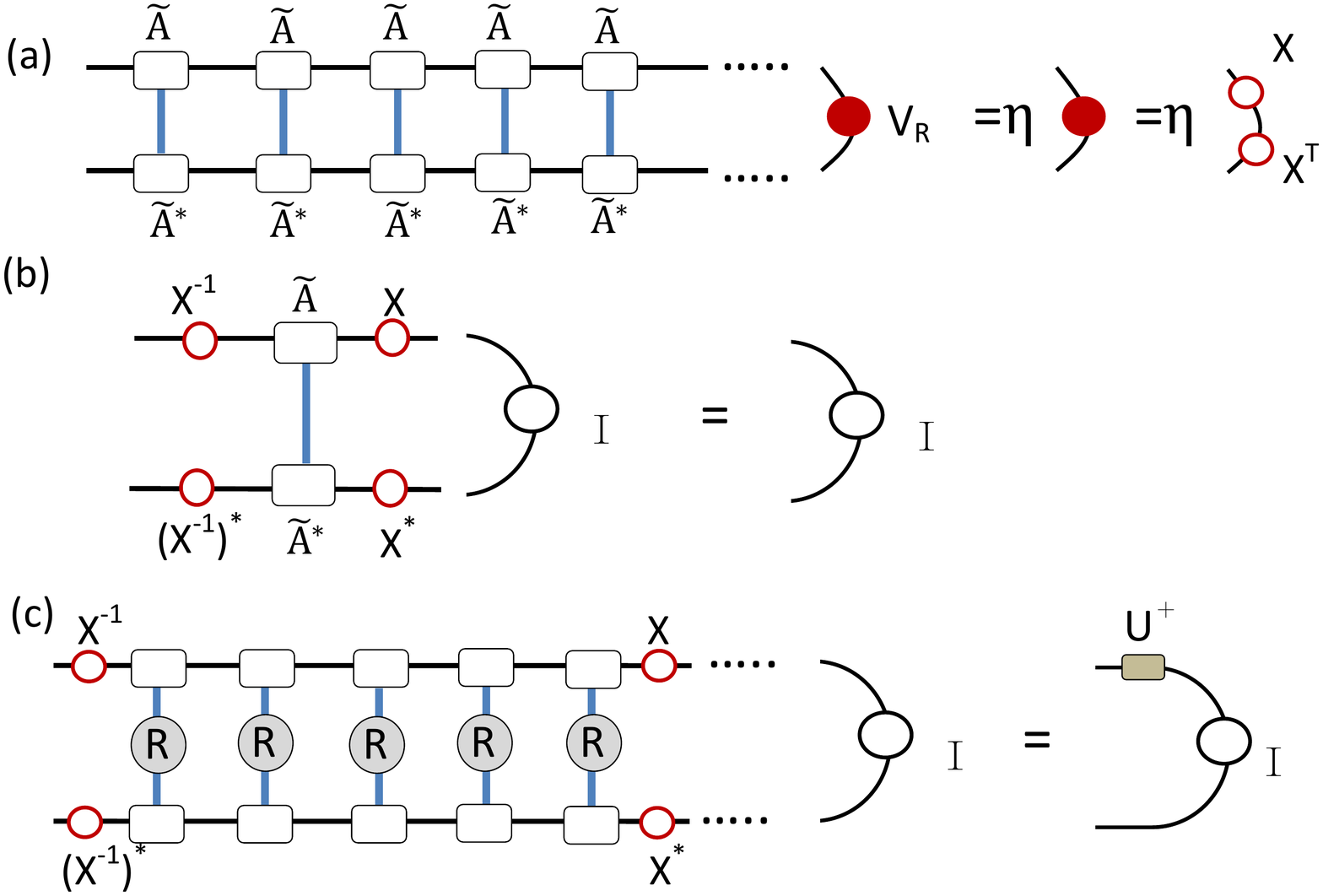,angle=0,width=8cm}}
  \caption[canonical]
  {(a) Diagrammatic representation of the right-hand side vector. (b) Condition of a MPS to be a canonical form. The transfer matrix have the identity as eigenvectors with eigenvalue $1$. (c) Insert operators to a TPS to find the projective representation.     }
   \label{canonical}
   \end{figure}

  \item  Determine  projective representations:

  To obtain the projective representations of symmetries, we need to insert symmetry operators $R$'s to the measured state.
  The generalized transfer matrix $\mathbb{G}$ is given by
  \begin{align}
  \mathbb{G}_{\hat{\alpha}\hat{\alpha}',\hat{\beta}\hat{\beta}'}=\sum_{SS'} \tilde{A}^{*S}_{\hat{\alpha}',\hat{\beta}'} R^{SS'} \tilde{A}^{S'}_{\hat{\alpha},\hat{\beta}},
  \end{align}
  where the operator $R^{SS'}$ are applied to the tensors.
  As $R$ is a symmetry operator, the generalized transfer matrix $\mathbb{G}$ has a largest eigenvalue $\eta$  of modulo $1$ as shown in Fig. \ref{canonical}c.
  Again we apply the generalized transfer matrix many times to an initial state, and the dominant state is related to a projective representations $U^a_g$.

\end{enumerate}

\section{Extracting the projective representations from exact diagonalization} \label{app_ED}

In this section, we will show how to obtain the projective representation via exact diagonalization.
For simplicity, we demonstrate this here only for SPTs.
First of all, we consider open boundary conditions for a Hamiltonian with $2N+1$ sites, and obtain the ground state wave function $|\Psi\rangle$ by using exact diagonalization.
Then, we want to determine the MPS representation of the middle site from $|\Psi\rangle$.
However, the degenerate ground states are likely to lead to ambiguities.

In particular, a 1D SPT phase has  generically degeneracies due to the edge modes at the ends of the chain.
For example, the $S=1$ antiferromagnetic chain in the Haldane phase has  spin $1/2$ edge states which lead to the degeneracy of ground state.   
We can add a small symmetry breaking field at the boundaries to split the edge state degeneracy and obtain an unique ground state $|\Psi\rangle$.

The procedure is covered in detail as follows.
We do a Schmidt decomposition of the bipartite splitting:  $\{ 1\ldots N-1 | N \ldots 2N+1\}$ of ground state $|\Psi\rangle$, so that
\begin{align}
 |\Psi\rangle=\sum_{i=1}^\chi \lambda_i|\alpha_i^{[1\ldots N-1 ]} \rangle_L |\alpha_i^{[N \ldots 2N+1 ]} \rangle_R,
\end{align}
where $|\alpha_i^{[1\ldots N-1 ]} \rangle_L $ and $|\alpha_i^{[N \ldots 2N+1 ]} \rangle_R$ form an orthogonal basis. Again we give a Schmidt decomposition of $|\Psi\rangle$ according to $[1\ldots N ]:[ N+1 \ldots 2N+1]$,
\begin{align}
 |\Psi\rangle=\sum_{i=1}^\chi \tilde{\lambda}_i|\beta_i^{[1\ldots N ]} \rangle_L |\beta_i^{[N+1 \ldots 2N+1 ]} \rangle_R.
\end{align}
After the above decompositions, the ground state can be expressed as
\begin{align}
 |\Psi\rangle=\sum_{i=1}^\chi A_{\alpha_i,\beta_i}^{s_N} |\alpha_i^{[1\ldots N-1 ]} \rangle_L |s_N\rangle |\beta_i^{[N+1 \ldots 2N+1 ]} \rangle_R,
\end{align}
where $s_N=1,\ldots,d_s$ with $d_s$ denoting the physical dimension and $\chi$ is the number of the Schmidt values. We then arrive at the standard representation of the MPS.
To have a true translationally invariant MPS, we have a fix the phase ambiguity of the virtual indices.
For inversion symmetric systems, this can be done by expressing for example the left Schmidt states as the inverted right ones.
It is easy to obtain the projective representation of the matrix $A_{\alpha,\beta}^{s}$ by following the procedures of Appendix \ref{app findU}.

 \end{document}